\documentclass[a4paper,11pt]{article}
\pdfoutput=1 

\usepackage{jinstpub} 

\usepackage{lineno}

\usepackage{enumerate}

\title{\boldmath Calibration of photomultiplier tubes} 

\author[a]{L. N. Kalousis} 

\affiliation[a]{Physics Department, National Technical University, 157 80 Zografou, Athens, Greece}

\emailAdd{leonidas.kalousis@gmail.com}

\abstract{The purpose of the present article is to demonstrate the calibration of photomultipliers with a gaussian single photoelectron response using a numerical method based on the Discrete Fourier Transform (DFT). 
Conventional techniques, commonly employed in the literature, use approximate models or brute force numerical calculations of the convolution integrals that lead to the charge response function of the photomultiplier, $S_R(x)$. 
In this publication, we explain how a truncated gaussian model for the single photoelectron amplification can lead to rigorous results if one leans on the DFT approach. 
The distinct feature of this procedure is that $S_R(x)$ is calculated to all orders in the Poisson mean $\mu$ that characterizes the light intensity and no approximations are needed. 
This scheme was applied to the calibration of the Hamamatsu R7081 photomultiplier tube and a comparison of the DFT approach with the more standard numerical integration method is also presented.
Last, toy Monte Carlo data were analyzed for different values of $\mu$ to understand the precision of the DFT method.}


\arxivnumber{2304.07292}

\begin{document}
\maketitle
\flushbottom

\section{Introduction}
\label{sec:intro}

Most photomultipliers have a single photoelectron response function that can be parameterized with a gaussian distribution. 
A prime example of this case is the ten inches Hamamatsu R7081 photomultiplier tube (PMT) model that has been used in several experiments in particle physics. 
Double Chooz, RENO and IceCube to name only but a few \cite{dc,reno,icecube}.
It is known to operate at a nominal gain of $10^7$ units of electron charge for about 1500~V of high voltage. 
It has a photocathode sensitivity that ranges between 300 and 600~nm with a maximum peak in quantum efficiency of roughly 25~$\%$ at about 400~nm. 
The good photoelectron resolution, the small dark noise, the low glass radioactivity levels and relatively low cost make R7081 an attractive choice for the instrumentation of various detectors. 

Several methods have been employed in the literature for the calibration of PMTs with a gaussian single photoelectron (SPE) response. 
Perhaps, the most rigorous has been presented in a seminal paper written by Dossi \emph{et al.}, ref.~\cite{dossi}. 
In the aforementioned publication, the SPE response was parametrized by a combination of an exponential distribution and a truncated gaussian. 
In our view, the key elements of the Dossi paper can be summarized in the following points:
\begin{enumerate}[i.]
\item the realization that the exponential term that models photoelectrons (PEs) that miss the first amplification stage is part of the signal (it adds to the gain) and,
\item the use of a properly normalized, truncated gaussian to describe the full amplification chain, avoiding thus the prediction of negative charges in the underlying formulae. 
\end{enumerate}
None of these points is entirely trivial. 
Unfortunately the final equation for the PMT charge response, $S_R(x)$, is rather complicated to be worked out analytically and one has to resort to some sort of an approximation. 
For example, the DarkSide collaboration solved the convolution integrals involved in the calculation of $S_R(x)$ numerically for the first two peaks, and the higher peaks were approximated by perfect gaussians~\cite{darkside}. 
That was adequate for the needs of DarkSide. 

In this work, we seek to apply the Dossi model to the calibration of the R7081 PMT using the numerical method first presented in ref.~\cite{me}. 
The distinct feature of this procedure is that $S_R(x)$ is calculated numerically to all orders in the Poissonian mean ($\mu$) and no approximations are needed. 
In section~\ref{sec:theo} we outline briefly the basic theory of gain determination and we present the Discrete Fourier Transform (DFT) technique exploited throughout the publication. 
In section~\ref{sec:data} we analyze data sets of R7081 PMT showcasing the validity of our study. 
A comparison with the common numerical integration method is included. 
In section~\ref{sec:mc} we present a Monte Carlo study showing the advantages of the DFT approach. 
We close this publication with some general remarks concerning gain calibration. 

\section{Standard theory}
\label{sec:theo}


Whenever a fixed number of photons is shot towards the photocathode of a PMT there is a certain probability that some will convert and create electrons (quantum efficiency). 
These PEs are then collected by the anodes and directed to the amplification chain with some certain probability (collection efficiency). 
The number of PEs ($n$) registered by the PMT is given by the well-known Poisson formula:
\begin{align}
P(n;\mu) = e^{-\mu}\frac{\mu^n}{n!}. \label{eq:poisson}
\end{align}
The Poisson mean, $\mu$, characterizes the light source and the quantum and collection efficiencies jointly. 
Now, if $S(x)$ is the probability density function (PDF) for a single PE to create a charge in the vicinity of $x$, 
the probability for $n$ PEs to produce $x$ is dictated by $S_n(x)$, where $S_n(x)$ is the $n$-times convolution of $S(x)$. 
The charge response of a PMT can be readily worked out:
\begin{align}
S_R(x) = &\sum_{n=0}^{+\infty} P(n;\mu)   S^{(n)}_R(x) \nonumber \\
            = & \sum_{n=0}^{+\infty} P(n;\mu) (S_n*B)(x). \label{eq:sr}
\end{align}
Note that a final convolution with the pedestal PDF, $B(x)$, is needed to include white noise from the electronics and other sources. 
Of course, $S_1(x)=S(x)$ and $S_0(x)=\delta(x)$. More details on the theory of PMT calibration can be found in ref.~\cite{me}.
The mean and standard deviation of $S_R(x)$, that is, $Q_R$ and $\sigma_R$ are equal to:
\begin{align}
Q_R  = & \ Q_0 + \mu Q_s \\
\sigma_R^2      = & \  \sigma_0^2 + \mu( \sigma_s^2 + Q_s^2 ),  
\end{align}
where $Q_0, \ \sigma_0$ are the mean and standard deviation of $B(x)$ respectively and $Q_s, \ \sigma_s$ those of $S(x)$.
The calculation of these formulae can be found in the appendix~\ref{appA}. 

The essence of the DFT approach to gain determination lies in the realization that the DFT of $S_R(x)$, $\tilde S_R(k)$, has a very simple formula~\cite{me}:
\begin{align}
\tilde S_R(k) = \tilde B(k) e^{ \mu( \tilde S(k) - 1 ) } , \label{eq:dft}
\end{align}
where $\tilde B(k), \ \tilde S(k)$ are the DFTs of $B(x)$ and $S(x)$ respectively. 
The fact that the series of eq.~\eqref{eq:sr} can be summed formally in the Fourier inverse space, 
is due to the form of the Poisson factors and the simple mathematical theorem that the DFT of $S_n(x)$ is $n$ powers of $\tilde S(k)$, $\tilde S^n(k)$. 
If one could invert eq.~\eqref{eq:dft} analytically then one has a formula of $S_R(x)$ in a closed form. In practice this is an impossibility.  
In order to progress, in ref.~\cite{me} we proposed to perform the forward DFT and inverse DFT calculations numerically using the fftw package~\cite{fftw}. 
For this purpose a C++/ROOT based software~\cite{root} was developed that calculates $S_R(x)$ for a given number of steps.
The code has been committed in a public github repository with several examples that can assist the interested reader~\cite{git}. 
We have analyzed all data in this article using this software.

\section{Data analysis}
\label{sec:data}

\subsection{Single photoelectron response model}

 The SPE response model of any given PMT can be parameterized by the general formula:
 \begin{align}
S(x) =    \ w \alpha e^{-\alpha x } H(x) + (1-w)g(x). \label{eq:S}
\end{align}
A few remarks are necessary here. 
First, the prefactor $w$ parameterizes, through the exponential distribution, the probability that a single PE will miss the first amplification stage. 
It ranges from zero to one. For $w=0$, all PEs are amplified according to the full chain. 
A better discussion of this can be found in ref.~\cite{dossi}. 
$H(x)$ is the Heaviside step function.  
The calculation of $Q_s$ and $\sigma_s$ is not at all trivial but not that difficult to perform. 
It is done in the appendix \ref{appB} and the result is:
\begin{align}
Q_s = & \ \frac{w}{\alpha} + (1-w)Q_g \label{eq:qs} \\ 
\sigma_s^2       = & \ \frac{w}{a^2} +(1-w)\sigma_g^2 + w(1-w)\left( Q_g -\frac{1}{a} \right)^2,
\end{align}
where $Q_g$ and $\sigma_g$ are the mean value and standard deviation of $g(x)$.
Note that when $w=0$, we have $Q_s=Q_g$ and $\sigma_s=\sigma_g$. 
On the other hand, for $w=1$ the mean value and standard deviation of $S(x)$ are those of the exponential term.  

 The $g(x)$ PDF model adopted by Dossi \emph{et al.} is~\cite{dossi}:
\begin{align}
g(x) =    \frac{1}{g_N} \frac{1}{\sqrt{2\pi}\sigma} e^{ - \frac{( x - Q )^2}{2\sigma^2}} \ H(x). \label{eq:g}
\end{align}
This characterizes the amplification of the full dynode system. 
It is a gaussian truncated at the negative values of $x$. 
The factor $g_N$ ensures that $g(x)$ is properly normalized and equals to: 
\begin{align}
g_N =  \frac{1}{2} \text{erfc} \left( -\frac{Q}{\sqrt{2}\sigma } \right).
\end{align}
erfc(x) is the complementary error function~\cite{error}. 
The mean value and standard deviation of $g(x)$ are deduced in the appendix~\ref{appC}. 
They are equal to:
\begin{align}
Q_g = & \  Q + \kappa       \label{eq:qg}         \\
\sigma_g^2       = & \  \sigma^2 - ( Q + \kappa )\kappa ,      
\end{align}
where $\kappa$ is given by the equation:
\begin{align}
 \kappa = \  \frac{1}{g_N} \frac{1}{\sqrt{2\pi}} \  \sigma \ e^{ - \frac{Q^2}{2\sigma^2}}. 
\end{align} 
One can see that as $Q$ increases, $\kappa$ approaches to zero due  to the damping exponential factor and $Q_g, \ \sigma_g$ approach $Q$ and $\sigma$ respectively.  

\subsection{Experimental data and results}

A PMT model based on eq.~\eqref{eq:S} and \eqref{eq:g} is quite difficult to derive in a closed form.
Even without the final convolution with $B(x)$, it is rather cumbersome to calculate $S_n(x)$ even for the lowest value of $n=2$. 
For example, for $n=2$ one has three terms to compute while for $n=3$ one has four different terms in $S_n(x)$ ! 
One has to resort to some sort of an approximation. 
For instance, in the original Dossi \emph{et al.} paper a simple formula was given for $S_R(x)$ where all $S_n(x)$ distributions above $n=1$ were modeled by symmetric gaussians. 
Additionally, in section~5 of that paper a brute force numerical method was presented. 
It is the purpose of this article to solve the Dossi model using the DFT procedure explained in section~\ref{sec:theo}. 

\begin{figure}[!t]
\centering
\includegraphics[width=9.5cm, height=6.4cm]{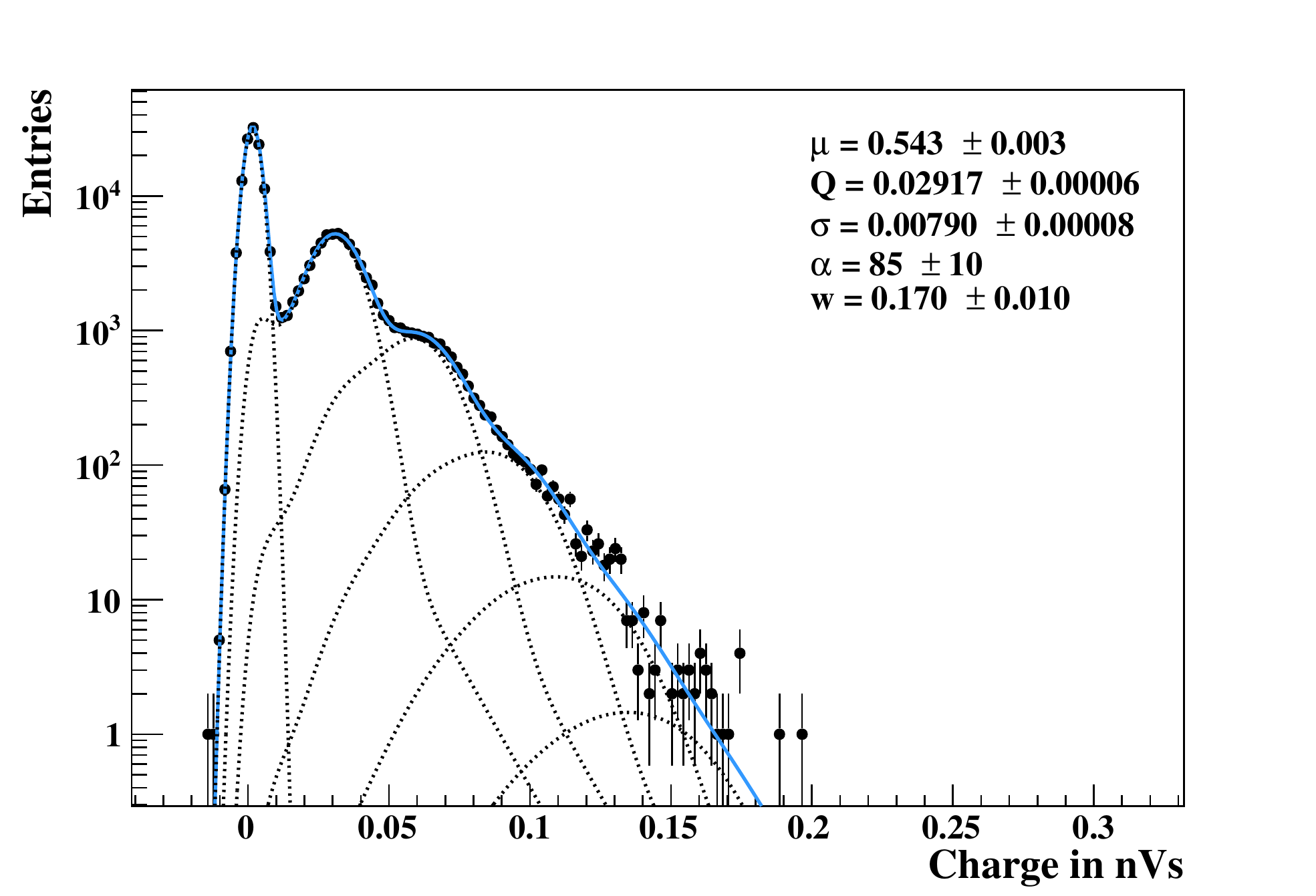} \\[1.5ex]
\includegraphics[width=9.5cm, height=6.4cm]{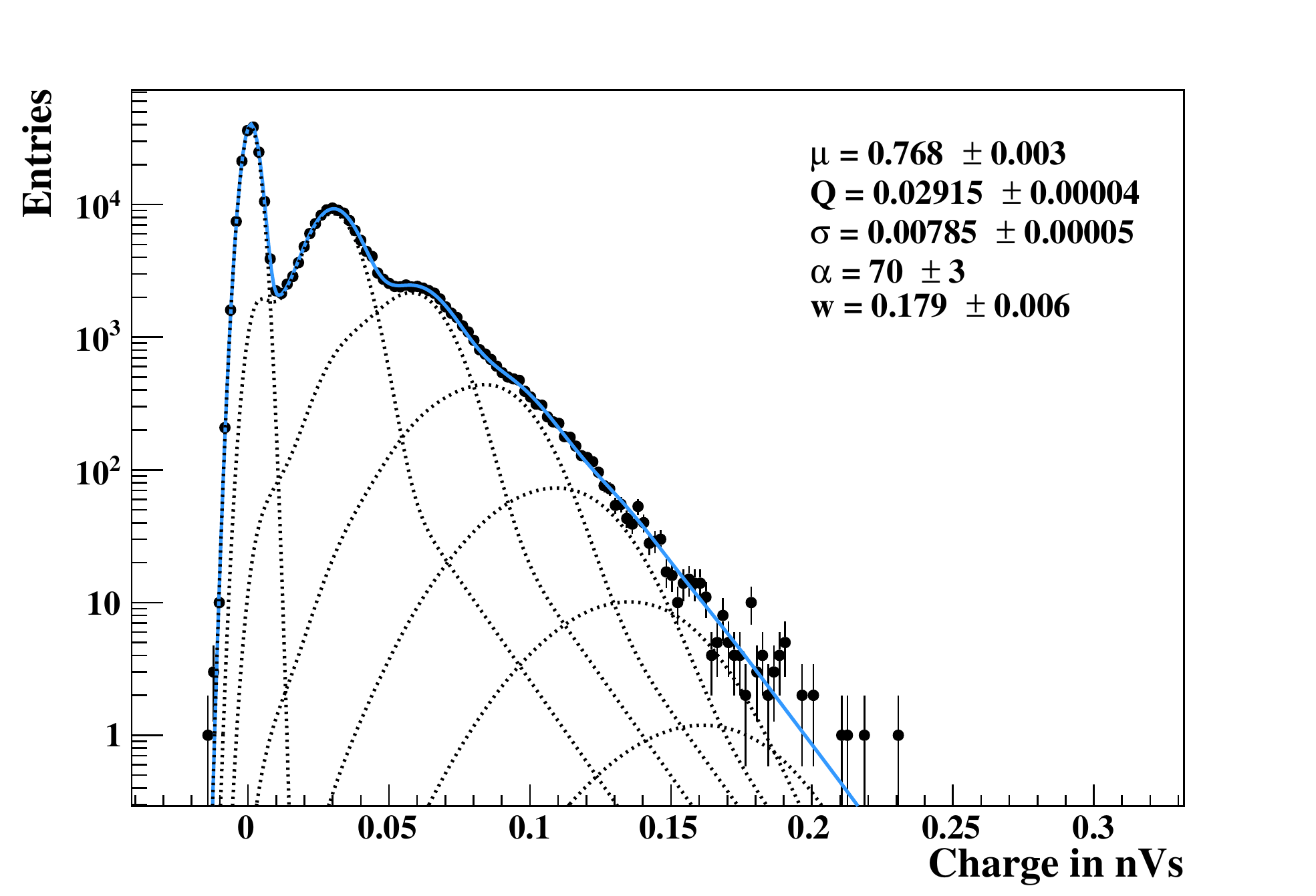}  \\[1.5ex] 
\includegraphics[width=9.5cm, height=6.4cm]{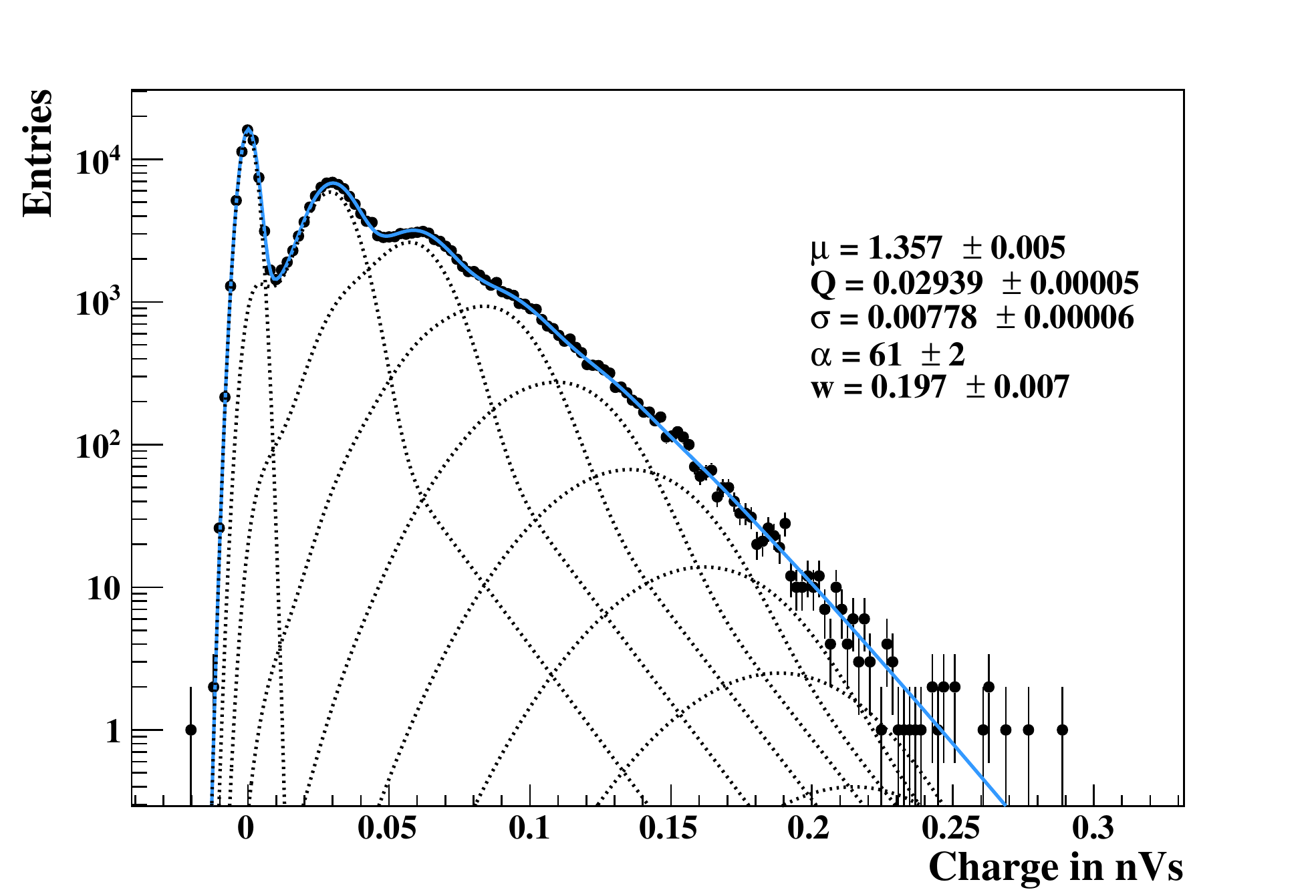} 
\caption{A few fits obtained using the Dossi \emph{et al.} model through the DFT method. 
The data are shown in the black dots and the best fit curve is shown in azure line. 
The dotted lines show the contributions of the various PE peaks.}
\label{fig:Q}
\end{figure}
Several data sets were taken with a R7081 PMT inside a light-tight box. 
The details of the experimental setup can be found in ref.~\cite{me}.
In general, the PMT was illuminated by an optical fiber connected to a light-emitting diode (LED).
The light pulses were produced at the LED by a fast pulse generator. 
The charge was readout by a LeCroy oscilloscope (WavePro 725Zi) triggering at the generator's second, duplicated channel. 
Like this, all pulses were recorded, including those with no PEs produced at the photocathode (pedestal). 
Figure~\ref{fig:Q} shows a few examples of the charge distributions obtained from our setup.  
To analyze the data and obtain the gain from the data points, we fitted the spectra with the $S_R(x)$ calculated using the DFT method.  
The procedure is best described in ref.~\cite{me}. 
First, a gaussian fit was performed on the pedestal alone to obtain approximate values for $Q_0$, $\sigma_0$ and $\mu$. 
The model was minimized using the Minuit2 software~\cite{Minuit2}. 

We should point out that for all the measurements included in this publication the optical fiber was placed at the center of the photocathode,
using a plastic halo-like structure, and that the fiber was always pointing vertically with respect to the surface of the PMT.\footnote{%
More details on the experimental setup, including pictures of the halo-like structure, can be found in ref.~\cite{thesis}.}
In this way one has to deal a single quantum efficiency (that at the center of the PMT) and we expect the assumption of eq.~\eqref{eq:poisson} to be quite valid. 
Note that the quantum efficiency of the PMT is expected to vary across the surface of the photocathode and a light source illuminating the whole surface of the PMT will not be 
accurately described by eq.~\eqref{eq:poisson}. In such cases one has to take into account the variance of the quantum efficiency across the incident angle and the Poissonian factors have to be modified. 
Note also that any possible bias in the extraction of the gain has to be attributed to the validity of the SPE response model (which was taken as an assumption)
and/or the stability of our setup. 

\begin{table}[t!]
\centering
\begin{tabular}{| c  || c | c | c | c || c |}
\hline
$\mu$                      & $w$                         &  $\alpha$         &  $Q$                               &  $\sigma$ & $\chi^2$/NDOF\\[0.6ex] \hline\hline
0.543 $\pm$ 0.003  & 0.170 $\pm$ 0.010  &  85 $\pm$ 10  & 0.02917 $\pm$ 0.00006 &  0.00790 $\pm$ 0.00008 & 1.52 \\
0.632 $\pm$ 0.003  & 0.180 $\pm$ 0.008  &  71 $\pm$   4  & 0.02920 $\pm$ 0.00006 &  0.00782 $\pm$ 0.00007 & 1.18 \\
0.768 $\pm$ 0.003  & 0.179 $\pm$ 0.006  &  70 $\pm$   3  & 0.02915 $\pm$ 0.00004 &  0.00785 $\pm$ 0.00005 & 1.60 \\
0.979 $\pm$ 0.003  & 0.196 $\pm$ 0.006  &  63 $\pm$   2  & 0.02923 $\pm$ 0.00005 &  0.00774 $\pm$ 0.00005 & 1.86 \\
1.357 $\pm$ 0.005  & 0.197 $\pm$ 0.007  &  61 $\pm$   2  & 0.02939 $\pm$ 0.00005 &  0.00778 $\pm$ 0.00006 & 1.68 \\
2.014 $\pm$ 0.008  & 0.196 $\pm$ 0.007  &  61 $\pm$   2  & 0.02935 $\pm$ 0.00006 &  0.00777 $\pm$ 0.00007 & 1.55
\\[0.6ex] \hline\hline
\end{tabular}
\caption{Summary of the R7081 PMT calibration results.}
\label{tab:money}
\end{table}

\begin{figure}[!t]
\centering
\includegraphics[width=12.3cm, height=8.8cm]{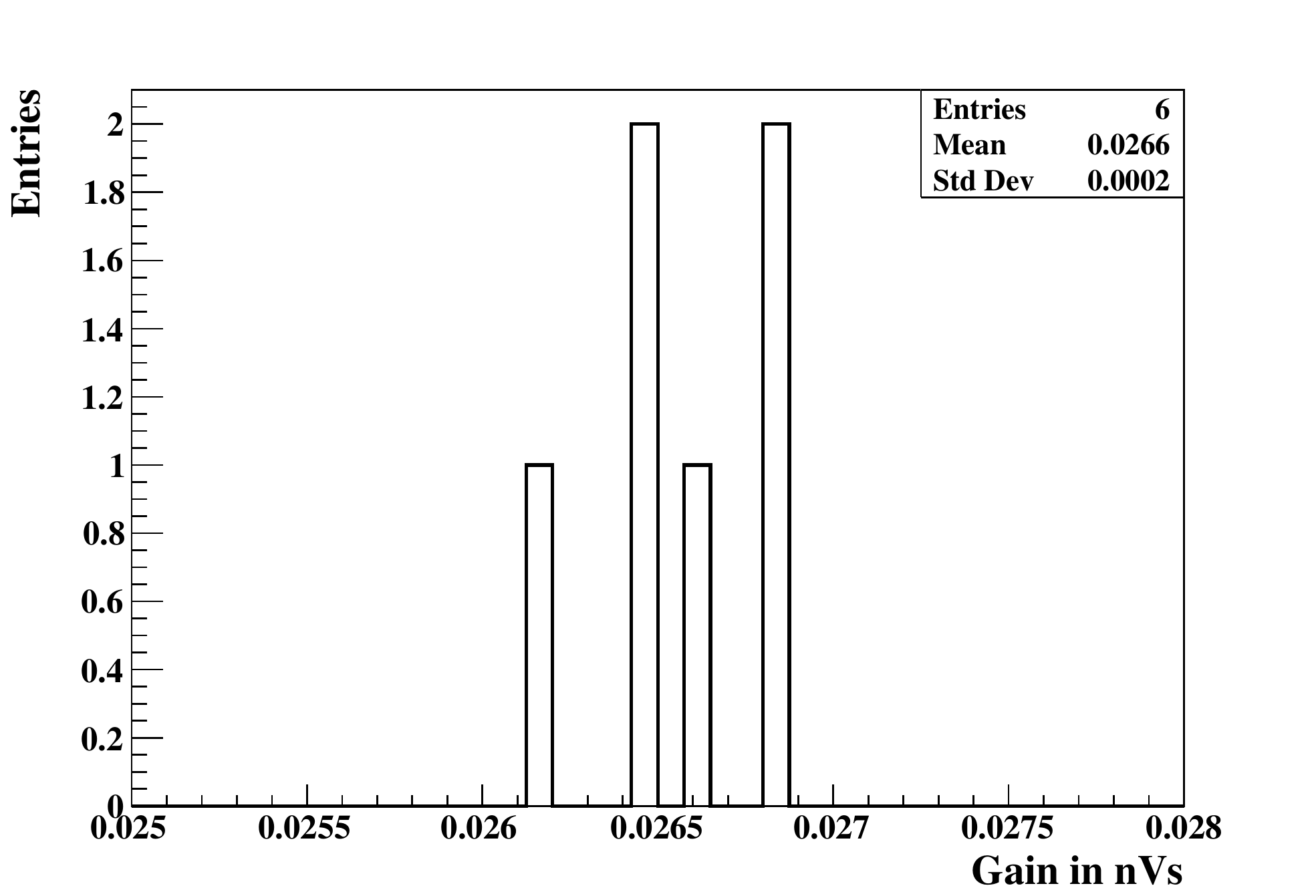} 
\caption{Distribution of the gain ($Q_s$) for the measurements compiled in table~\ref{tab:money}. }
\label{fig:g}
\end{figure}

Figure~\ref{fig:Q} shows the best fit curves in azure line. 
One can readily see that the model follows closely the data for all examples ($\chi^2$/NDOF was always close to one). 
Furthermore, more data were taken with increasing light intensity to assess the stability of gain determination. 
The results are gathered in table~\ref{tab:money}. 
Even though strong correlations exist between the various parameters, and even though there's a drift in the extraction of $w$ and $\alpha$  in particular,
the gain as calculated by the eq.~\eqref{eq:qs} and \eqref{eq:qg} is quite stable inside the $\mu$~$\sim$~0.5 -- 2.0 PE plateau. 
Figure~\ref{fig:g} shows the gain distribution for the results of table~\ref{tab:money}. One readily sees that $Q_s$ is stable within a $\sim$~1~\% range. 
Two further remarks should be made. First, one should never identify gain with $Q_g$ as this will overestimate the true gain. 
On the contrary, the gain should be calculated by the weighted average of eq.~\eqref{eq:qs}. 
Second, in the particular case of the R7081 PMT the truncation of the gaussian PDF at negative charge values has little impact on the final results; 
$Q_g$ is very close to $Q$. 

\subsection{Comparison with the numerical integration method}

To assess the advantages of the DFT approach towards gain determination, we analyzed the same R7081 data set using the numerical method presented in ref.~\cite{dossi}.   
Note that according to this procedure all the integrals of the convolutions in eq.~\eqref{eq:sr} are calculated numerically. 
In practice, this was implemented in our analysis software using multiple for loops. 
Unfortunately, the execution time to compute several $S_n(x)$ distributions becomes increasingly large, and thus, is rendering this method almost inapplicable. 
To evade this difficulty we calculated the first PE peak\footnote{That is ($S*B$)(x).} numerically and the higher PE peaks were approximated with symmetric gaussians. 

\begin{table}[t!]
\centering
\begin{tabular}{| c  | c  | c | c | }
\hline  \multicolumn{2}{|c|}{ DFT approach} & \multicolumn{2}{|c|}{Numerical integration} \\[0.6ex] \hline\hline
  Gain, $Q_s$ & $\chi^2$/NDOF & Gain, $Q_s$  & $\chi^2$/NDOF \\ \hline
 0.0262 & 1.52 & 0.0262 & 2.33 \\
 0.0265 & 1.18 & 0.0263 & 2.03 \\ 
 0.0265 & 1.60 & 0.0264 & 2.97 \\ 
 0.0266 & 1.86 & 0.0264 & 3.31 \\ 
 0.0268 & 1.68 & 0.0266 & 2.85 \\
 0.0268 & 1.55 & 0.0267 & 2.96 
\\[0.6ex] \hline\hline 
\multicolumn{2}{|c|}{ 1 sec } & \multicolumn{2}{|c|}{ 11 sec } \\  \hline
\end{tabular}
\caption{Comparison of the DFT approach and the numerical integration method.} 
\label{tab:comp}
\end{table}

Table~\ref{tab:comp} shows the results of this study.
For comparison purposes the figures from the DFT analysis are also included. 
First columns show the gain ($Q_s$) obtained from the minimization procedure and second columns show the $\chi^2$/NDOF. 
The last row shows the execution time that each method took for the analysis of the dataset presented in table~\ref{tab:money}.
From these numbers the following conclusions can be drawn:
\begin{enumerate}[i.]
\item The first thing that can be observed from table~\ref{tab:comp} is that the two methods give almost identical figures insofar as gain determination is concerned. 
Inside the $\mu\sim$ 0.5 -- 2.0 plateau both techniques return the same gain. 
\item Second, the DFT approach gives a better fit, since the $\chi^2$/NDOF is systematically smaller than that of the numerical method. 
\item Last, the DFT method is almost ten times faster than the numerical integration. 
\end{enumerate}
The data that we took and analyzed in this article lie well inside the $\mu\sim$ 0.5 -- 2.0 window. 
Part of this decision was motivated by the fact that a pronounced pedestal always assist the fitter and enhances gain determination through strong constraints in $Q_0$, $\sigma_0$ and $\mu$. 
To understand how our method performs for large $\mu$, in the next section we analyzed sets of simulated data.  

\section{Simulated data}
\label{sec:mc}

Series of several toy Monte Carlo SPE spectra were generated. 
The algorithm that was used to produce these data sets can be summarized in the following steps:
\begin{enumerate}[i.]
\item First, a number of PEs was thrown from a Poisson distribution of mean value $\mu$.
\item For each PE, a charge was picked randomly from the SPE distribution of eq.~\eqref{eq:S} and \eqref{eq:g}. 
The total charge was calculated by summing the individual charges that each PE deposits. 
\item A charge was thrown from the gaussian distribution of the pedestal and summed to the total charge obtained in step two. 
\item The final, total charge was filled in a histogram.  
\item This procedure was repeated $2.5\  10^6$ times and a histogram of $2.5\  10^6$ entries was produced. 
\end{enumerate}
The parameters of the pedestal and SPE distributions were set to match those of the first line in table~\ref{tab:money}. 
For each value of $\mu$ one hundred toy spectra were generated and fitted with both the DFT and the numerical integration methods. 
Data were generated for $\mu$ inside the 0.5 -- 5.0 range. For each set of one hundred toys, the distribution of the relative deviation from the true gain ($Q_s$):
\begin{align}
\Delta Q_s = \frac{Q_s^\prime-Q_s}{Q_s} ,
\end{align}  
was plotted. We note that $Q_s^\prime$ is the gain parameter returned by the minimizer. 

\begin{figure}[!t]
\centering
\includegraphics[width=12.4cm, height=8.9cm]{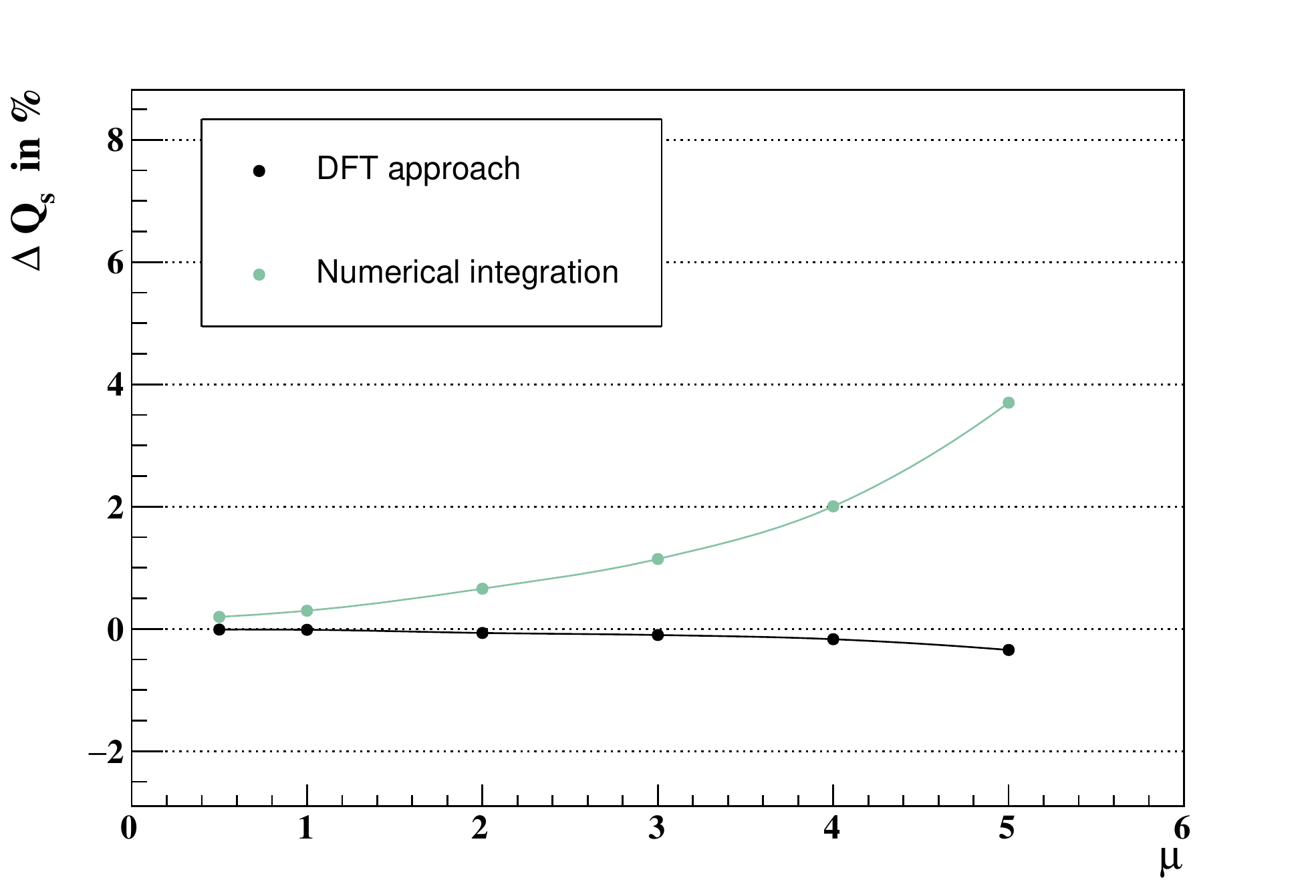} 
\caption{Comparison of the DFT (black) and numerical integration (green) methods. }
\label{fig:comp}
\end{figure}

Figure~\ref{fig:comp} shows the results of this exercise. 
The $x$ axis shows $\mu$ and the $y$ axis shows the mean value of the $\Delta Q_s$ distribution for the one hundred toys. 
The black dots depict the results from the DFT approach and the green dots those of the numerical integration respectively. 
One sees that DFT provides excellent results inside the $\mu = $ 0.5 -- 5.0 range and a deviation of $\sim$ 0.3\% is observed only at the high value of $\mu=5$. 
In contrast, the numerical integration method starts to deviate significantly from the true gain and for $\mu=5$ the deviation reaches the value of $\sim$ 4~\%.

\section{Outlook}
\label{sec:outro}

In this article, we presented the calibration of PMTs with gaussian SPE charge response. 
The analysis was based on a truncated gaussian PDF for the SPE response and we relied on the DFT method to solve the $S_R(x)$ model numerically. 
We have showed that despite the complications involved in the multidimensional fit of the data, one can extract the gain with good precision. 
In particular, for the R7081 PMT the gain can be determined with $\sim$~1~\% accuracy or better within the $\mu$~$\sim$ 0.5 -- 2.0 plateau. 
Attention was paid to emphasize the fact that the gain should not be confused with the mean value of the SPE response $g(x)$ and, instead, a weighted average between the exponential and gaussian terms should be preferred. 
The same procedure can be applied to other PMTs which share the same characteristics in charge response with the R7081 PMT. 
The analysis software used in these studies exists in a public github repository and can be used by other investigators \emph{mutatis mutandis}. 

A comparison of the DFT and numerical integration methods was also attempted. 
In particular we showed, using R7081 data, that within the 0.5 -- 2.0 PE range both techniques provide consistent results, but DFT gives better $\chi^2$/NDOF and runs much faster. 
In this respect, we should point out that DFT is more appropriate for the analysis of large data samples. 
Several analysis were performed with toy Monte Carlo data for various values of $\mu$. 
Again, it was demonstrated that the DFT approach outperforms the numerical integration having an accuracy of better than 0.5~\% inside the $\mu\sim $ 0.5 -- 5.0 window. 

The following remarks are necessary. 
In this publication we have only tried to demonstrate the calibration of the R7081 PMT model in just the simplest case. 
That is, when a single PMT was placed inside a dark, light-tight box and with the optical fiber positioned at the center of the photocathode, and always pointing vertically towards the surface of the PMT. 
We have demonstrated that in this simple example, the gain remains remarkably stable (within $\sim 1\%$ or better) inside the $\mu \sim 0.5 - 2.0$ plateau. 
We should note that in large (monolithic) detectors equipped with a sizable number of PMTs this simplistic picture ceases to apply. 
In particular, in those circumstances we can expect the extraction of the gain to depend on the geometry of the detector and the position of the event. 
 We can only expect that the accuracy achieved in this article will not be attainable in such cases.
The question of the \emph{in situ} gain calibration of similar, complicated instruments lies beyond the scope of this article and was not treated here. 
More details on the energy and spatial resolution of large-volume liquid scintillator detectors can be found elsewhere~\cite{smirnov}.

\appendix

\section{Calculation of $Q_R$ and $\sigma_R$}
\label{appA}

In order to compute $Q_R$ and $\sigma_R$ we first write down the formulae:
\begin{align}
\sum_{n=0}^{+\infty} P( n; \mu ) & = 1, \\
\sum_{n=0}^{+\infty} n P( n; \mu ) & = \mu, \\
\sum_{n=0}^{+\infty} n^2 P( n; \mu ) & = \mu ( \mu + 1 ). 
\end{align} 
The first equation stems from probability conservation and it is very easy to derive.
The other two can be proved by a shifting of the summing parameter $n$. 
We also note the formulae which are deduced from the properties of the convolution:
\begin{align}
< ( S_n * B )(x) > \ = \ & Q_0 + n Q_s, \\
\text{Var}[ ( S_n * B )(x) ] \ = \ & \sigma_0^2 + n\sigma_s^2. 
\end{align} 
We simplify our notation by setting $S_R^{(n)}(x)=( S_n * B )(x)$. 
Using these equations, the mean value $Q_R$ becomes:
\begin{align}
Q_R \ = \ & < S_R (x) >  \nonumber \\
 \ = \ & \sum_{n=0}^{+\infty} P( n; \mu ) \int_{-\infty}^{+\infty} x S_R^{(n)}(x) \ dx \nonumber \\
  \ = \ & \sum_{n=0}^{+\infty} P( n; \mu ) ( Q_0 + n Q_s ) \nonumber \\
  \ = \ & Q_0 + \mu Q_s.   
\end{align} 
On the other hand, to calculate the variance $\sigma_R^2$ we first find the integral:
\begin{align}
\int_{-\infty}^{+\infty} x^2 S_R(x) \ dx \ = & \  \sum_{n=0}^{+\infty} P( n; \mu )   \int_{-\infty}^{+\infty} x^2 S_R^{(n)}(x) \ dx \nonumber \\
\ = & \ \sum_{n=0}^{+\infty} P( n; \mu ) ( \sigma_0^2 + n\sigma_s^2 + ( Q_0 +n Q_s )^2 ) \nonumber \\
\ = & \ \sum_{n=0}^{+\infty} P( n; \mu ) ( \sigma_0^2 + n\sigma_s^2 + Q_0^2  + n^2 Q_s^2  + 2 n Q_0 Q_s ) \nonumber \\
\ = & \ \sigma_0^2 + \mu\sigma_s^2 + Q_0^2 + \mu ( \mu + 1 ) Q_s^2 + 2 \mu Q_0 Q_s \nonumber \\
\ = & \ \sigma_0^2 + \mu\sigma_s^2 + \mu Q_s^2 + ( Q_0 + \mu Q_s )^2. 
\end{align}
Where in the forth line we made use of the identities provided in the beginning of this section. 
The variance now becomes:
\begin{align}
\sigma_R^2 \ = \ & \text{Var}[ S_R(x)] \nonumber \\
\ = \ & \int_{-\infty}^{+\infty} x^2 S_R(x) \ dx - Q_R^2 \nonumber \\
\ = \ & \sigma_0^2 + \mu( \sigma_s^2 + Q_s^2 ).
\end{align}

\section{Variance of $S(x)$}
\label{appB}

The mean value of $S(x)$ is straightforward to work out and its proof will not be presented here. \ \ \ 
We only remark that the mean of the sum of two terms equals the sum of the two individual means. 
\begin{align}
Q_s =  \frac{w}{\alpha} + (1-w)Q_g 
\end{align}
The variance on the other hand is quite complicated and it will be treated in great detail. 
We first remind the reader that:
\begin{align}
\sigma^2_s \  =  \ & \int_{-\infty}^{+\infty} x^2 S(x) \ dx - Q_s^2, \label{eq:ss}
\end{align}
and we proceed to calculate each term in eq.~\eqref{eq:ss} separately. 
\begin{align}
Q_s \ = \ &  \left(\  \frac{w}{\alpha} + (1-w)Q_g  \ \right)^2 \nonumber \\
\ = \ & \frac{w^2}{\alpha^2} + 2 \frac{w(1-w)}{\alpha} Q_g + (1-w)^2 Q_g^2. \label{eq:e1}
\end{align}
\begin{align}
\int_{-\infty}^{+\infty} x^2 S(x) \ dx  \  =  \ & w   \int_{0}^{+\infty} x^2 \alpha e^{-\alpha x } \ dx + (1-w)\int_{-\infty}^{+\infty} x^2 g(x) \ dx  \nonumber \\
 \  =  \ & \frac{2w}{\alpha^2} + (1-w)( \sigma_g^2 + Q_g^2 ) \label{eq:e2}
\end{align}
Plugging eq.~\eqref{eq:e1} and \eqref{eq:e2} into $\sigma_R$ we have:
\begin{align}
\sigma^2_s \  =  \ & \frac{2w}{\alpha^2} + (1-w)( \sigma_g^2 + Q_g^2 )   -  \frac{w^2}{\alpha^2} - 2 \frac{w(1-w)}{\alpha} Q_g -  (1-w)^2 Q_g^2 \nonumber \\
                \  =  \ & \frac{w}{\alpha^2} + (1-w) \sigma_g^2  + \frac{w(1-w)}{\alpha^2} + w(1-w)Q_g^2 - 2 \frac{w(1-w)}{\alpha} Q_g \nonumber \\
                \  =  \ & \frac{w}{\alpha^2} + (1-w) \sigma_g^2  +  w(1-w)\left( \  \frac{1}{\alpha^2} +Q_g^2 - 2 \frac{Q_g}{\alpha} \  \right) \nonumber \\
                 \  =  \ & \frac{w}{\alpha^2} + (1-w) \sigma_g^2  + w(1-w)\left( \ Q_g - \frac{1}{\alpha}  \ \right)^2.
\end{align}

\section{Mean value and variance of $g(x)$}
\label{appC}

%

\subsection*{Mean value}

The mean value of $g(x)$ is equal to:
\begin{align}
Q_g \  = \ & \int_{-\infty}^{+\infty} x g(x) \ dx \nonumber \\
\  = \ &  \frac{1}{g_N}\frac{1}{\sqrt{2\pi}\sigma} \int_{0}^{+\infty} x e^{-\frac{(x-Q)^2}{2\sigma^2}} \ dx. \label{eq:intqg}
\end{align}
The integral of eq.~\eqref{eq:intqg} can be calculated using the substitution of variables:
\begin{align}
u \  = \ \frac{x-Q}{\sqrt{2}\sigma}.
\end{align}
It turns out to be:
\begin{align}
I \  = \ & \int_{0}^{+\infty} x e^{-\frac{(x-Q)^2}{2\sigma^2}} \ dx \nonumber \\
\  = \ & \int_{  -\frac{Q}{\sqrt{2}\sigma}.  }^{+\infty}  ( \sqrt{2}\sigma u + Q ) e^{-u^2} \sqrt{2}\sigma \ du \nonumber \\
\  = \ & 2 \sigma^2 \int_{  -\frac{Q}{\sqrt{2}\sigma} }^{+\infty}  u  e^{-u^2}  dx + \sqrt{2}\sigma Q \int _{  -\frac{Q}{\sqrt{2}\sigma}.  }^{+\infty} e^{-u^2}  dx \nonumber \\
\  = \ & \sigma^2 e^{-  \frac{Q^2}{2\sigma^2 } } + \sqrt{2\pi}\sigma Q \ \frac{1}{2}\left( \ 1 - \text{erf}\left( -\frac{Q}{\sqrt{2}\sigma} \right) \ \right) \nonumber \\
\  = \ & \sigma^2 e^{-  \frac{Q^2}{2\sigma^2 } } + \sqrt{2\pi}\sigma Q \  \frac{1}{2} \text{erfc}\left( -\frac{Q}{\sqrt{2}\sigma} \right)  \nonumber \\
\  = \ & \sigma^2 e^{-  \frac{Q^2}{2\sigma^2 } } + \sqrt{2\pi}\sigma Q \ g_N.
\end{align}
If we substitute $I$ into eq.~\eqref{eq:intqg} we then have:
\begin{align}
Q_g \  = Q +\frac{1}{g_N}\frac{\sigma}{\sqrt{2\pi}} e^{-  \frac{Q^2}{2\sigma^2 } }. 
\end{align}
This equation can be simplified by setting: 
\begin{align}
\kappa  = \  \frac{1}{g_N}\frac{\sigma}{\sqrt{2\pi}} e^{-  \frac{Q^2}{2\sigma^2 } },
\end{align}
so that $Q_g$ is equal to:
\begin{align}
Q_g \  = Q + \kappa. 
\end{align}

\subsection*{Variance}

To calculate the variance of $g(x)$ we proceed to find the integral:
\begin{align}
 \int_{-\infty}^{+\infty} x^2 g(x) \ dx \ = \ \frac{1}{g_N} \frac{1}{\sqrt{2\pi}\sigma} \int_{0}^{+\infty} x^2 e^{-\frac{(x-Q)^2}{2\sigma^2}} \ dx.
\end{align}
\begin{align}
 I \ =  \ &\int_{0}^{+\infty} x^2 e^{-\frac{(x-Q)^2}{2\sigma^2}} \ dx  \nonumber \\
 \  = \ & \int_{  -\frac{Q}{\sqrt{2}\sigma}.  }^{+\infty}  ( \sqrt{2}\sigma u + Q )^2 e^{-u^2} \sqrt{2}\sigma \ du \nonumber \\
 \  = \ & \sqrt{2\pi}\sigma g_N ( \sigma^2 + Q^2 ) + \sigma^2 Q e^{-  \frac{Q^2}{2\sigma^2 } }
\end{align}
\begin{align}
 \int_{0}^{+\infty} x^2 g(x) \ dx \ = \ &  \frac{1}{g_N} \frac{1}{\sqrt{2\pi}\sigma} I \ \nonumber \\
 \  = \ & \sigma^2 + Q^2 +  \frac{1}{g_N}\frac{\sigma Q }{\sqrt{2\pi}} e^{-  \frac{Q^2}{2\sigma^2 } }  \nonumber \\
  \  = \ & \sigma^2 + Q^2 + \kappa Q
\end{align}
Finally, the variance $\sigma_g^2$ is given by the formula:
\begin{align}
\sigma_g^2 \ = \ & \int_{-\infty}^{+\infty} x^2 g(x) \ dx - Q_g^2 \nonumber \\
\  = \ & \sigma^2 + Q^2 + \kappa Q - ( Q+\kappa)^2 \nonumber \\
\  = \ & \sigma^2  - ( Q + \kappa )\kappa. 
 \end{align}

\acknowledgments

The data used in this publication were taken in the laboratory of M. Dracos and we wish to thank him for allowing us to use them for the purposes of this communication. 



\end{document}